\documentclass[a4paper,fleqn,usenatbib,useAMS]{mnras}
\usepackage{graphicx}	% Including figure files
\usepackage{amsmath}	% Advanced maths commands
\usepackage{amssymb}	% Extra maths symbols
\usepackage{multicol}        % Multi-column entries in tables
\usepackage{bm}		% Bold maths symbols, including upright Greek
\usepackage{pdflscape}	% Landscape pages
\usepackage{natbib}
\usepackage{comment}
\title[Superflares on late-M dwarfs]{$K2$ Ultracool Dwarfs Survey. V. High superflare rates on rapidly rotating late-M dwarfs}

\author[R. R. Paudel et al.]{R. R. Paudel$^{1}$\thanks{Contact e-mail: \href{mailto:rpaudel@udel.edu}{rpaudel@udel.edu}},
 J. E. Gizis$^{1}$, 
 D. J. Mullan$^{1}$, 
 S. J. Schmidt$^{2}$, 
 A. J. Burgasser$^{3}$, 
\newauthor{P. K. G. Williams$^{4}$, 
A. Youngblood$^{5}$}
\\
$^{1}$Department of Physics and Astronomy, University of Delaware, Newark, DE, 19716, USA\\
$^{2}$Leibniz-Institute for Astrophysics Potsdam (AIP), An der Sternwarte 16, 14482, Potsdam, Germany \\
$^{3}$Center for Astrophysics and Space Science, University of California San Diego, La Jolla, CA 92093, USA\\
$^{4}$Harvard-Smithsonian Center for Astrophysics, 60 Garden Street, Cambridge, MA 02138, USA \\
$^{5}$NASA Goddard Space Flight Center, Greenbelt, MD 20771, USA}
\pubyear{2018}

\begin{document}
\label{firstpage}
\pagerange{\pageref{firstpage}--\pageref{lastpage}}
\maketitle

\begin{abstract}
We observed strong superflares (defined as flares with energy in excess of 10$^{33}$ erg) on three late-M dwarfs: 2MASS J08315742+2042213 (hereafter 2M0831+2042; M7 V), 2MASS J08371832+2050349 (hereafter 2M0837+2050; M8 V) and 2MASS J08312608+2244586 (hereafter 2M0831+2244; M9 V). 2M0831+2042 and 2M0837+2050 are members of the young ($\sim$700 Myr) open cluster Praesepe. The strong superflare on 2M0831+2042 has an equivalent duration (ED) of 13.7 hr and an estimated energy of 1.3 $\times$ 10$^{35}$ erg. We observed five superflares on 2M0837+2050, on which the strongest superflare has an ED of 46.4 hr and an estimated energy of 3.5 $\times$ 10$^{35}$ erg. This energy is larger by 2.7 orders of magnitude than the largest flare observed on the older (7.6 Gyr) planet-hosting M8 dwarf TRAPPIST-1. Furthermore, we also observed five superflares on 2M0831+2244 which is probably a field star. The estimated energy of the strongest superflare on 2M0831+2244 is 6.1 $\times$ 10$^{34}$ erg. 2M0831+2042, 2M0837+2050 and 2MASS J0831+2244 have rotation periods of 0.556$\pm$0.002, 0.193$\pm$0.000 and 0.292$\pm$0.001 d respectively, which we measured by using $K2$ light curves. We compare the flares of younger targets with those of TRAPPIST-1 and discuss the possible impacts of such flares on planets in the habitable zone of late-M dwarfs.
\end{abstract}

\begin{keywords}
stars: individual: (2MASS J08371832+2050349, 2MASS J08315742+2042213, 2MASS J08312608+2244586, TRAPPIST-1) - stars: activity - stars: low-mass - stars: flare
\end{keywords}

%%%%%%%%%%%%%%%%%%%%%%%%%%%%%%%%%%%%%%%%%%%%%%%%%%

%%%%%%%%%%%%%%%%% BODY OF PAPER %%%%%%%%%%%%%%%%%%

% The MNRAS class isn't designed to include a table of contents, but for this document one is useful.
% I therefore have to do some kludging to make it work without masses of blank space.
%\begingroup
%\let\clearpage\relax
%\tableofcontents
%\endgroup
\newpage

\section{Introduction}
M dwarfs make up $\sim$75\% of all main sequence stars in the local stellar population and are commonly referred to as red dwarfs \citep{2017AJ....154..124C}. They are smaller, cooler and less luminous than the Sun. Using the full four-year $Kepler$ data, \cite{2015ApJ...807...45D} estimated an occurence rate of 0.24$^{+0.18}_{-0.08}$ Earth-size planets and 0.21$^{+0.11}_{-0.06}$ super-Earths per M dwarf habitable zone (HZ). These results and the discovery of the TRAPPIST-1 planetary system \citep{2016Natur.533..221G,2017Natur.542..456G,2017NatAs...1E.129L} demonstrate that there is a significant chance of finding habitable planets around M dwarfs. The \textit{Transiting Exoplanet Survey Satellite (TESS; \citealt{2014JAVSO..42..234R})} will find many more HZ planets orbiting M dwarfs \citep{2018ApJS..239....2B}. With the discovery of many such planets, an essential next step in exoplanet research is to identify those with maximum probability of being habitable, especially those bright enough to be characterized by upcoming missions such as NASA's \textit{James Webb Space Telescope (JWST)}. \\
\\
Because the M dwarfs have smaller luminosity $L$ than the sun, in order to have the right temperature for liquid water to exist, the HZ around them must have average orbital radii $R$ ($\sim L^{0.5}$) that are a fraction of an AU. TRAPPIST-1, with $L$ = 5 $\times$ 10$^{-4}$  $L_{\odot}$ \citep{2018ApJ...853...30V} has an HZ at 0.022 AU, i.e. closer to the parent star by a factor of $\sim$50 than the Earth is to the Sun.
Furthermore, M dwarfs are typically magnetically active, with frequent flares from systems with ages of a few million years to billions of years \citep{1976ApJS...30...85L,2016ApJ...829...23D,2018ApJ...858...55P}. CFHT-BD-Tau 4 is an example of a very young $\sim$1 Myr old M7 dwarf on which a superflare with energy up to 10$^{38}$ erg, was observed \citep{2018ApJ...861...76P} and TRAPPIST-1 is an example of an old star with age 7.6$\pm$2.2 Gyr \citep{2017ApJ...845..110B} on which multiple flares were observed \citep{2018ApJ...858...55P}. The energy of the strongest flare observed on TRAPPIST-1 was 7.1$\times$10$^{32}$ erg in the ultra-violet/visible/infra-red wavelengths \citep{2018ApJ...858...55P}. SDSS J022116.84+194020.4 (M8 dwarf) is another example of a late-M dwarf, whose estimated age is in between $\sim$200 Myr and few Gyr, and on which a superflare with a total $U$-band energy of $\sim$10$^{34}$ erg was observed \citep{2014ApJ...781L..24S}. The closeness of the HZ to the parent star in an M dwarf planetary system implies that any planets located in the HZ of M dwarfs are more likely to be exposed to enhanced X-rays, UV radiation, high energy particles and coronal mass ejections (CMEs) associated with flares on the parent star. For example, if a flare  occurs on Proxima Centauri (Prox Cen; M5.5 dwarf) with the same energy as a typical solar flare, then the HZ planet Prox Cen b receives 250$\times$ more X-rays than Earth, 30$\times$ more EUV flux and 10$\times$ more FUV flux \citep{2016A&A...596A.111R}. The intense radiation and energetic particles may perturb the thermo-chemical equilibrium of the planet's atmosphere, including the destruction of ozone layer and loss of surface water.\\ \\
\cite{2010AsBio..10..751S} and 
\cite{2017arXiv171108484T} have reported on modeling the effects of flares on planetary atmospheres. \cite{2010AsBio..10..751S} studied the possible impacts of the 1985 April 12 flare from the dM3 star AD Leo \citep{1991ApJ...378..725H}, on an Earth-like planet in the HZ of this mid-M dwarf. Likewise, \cite{2017arXiv171108484T} studied the effects of high flare rate and high flare energies (10$^{30.5}$ - 10$^{34}$ erg) of the dM4 flare star GJ1243 on an Earth-like planet. In general, both studies find that if the flare output consists only of photons, then no significant ozone layer destruction occurs. In order to destroy the ozone, the main contributors must be energetic particles analogous to the solar energetic particles (SEP) which are generated along with coronal mass ejections (CMEs) in large solar flares. Assuming that particle fluxes can be generated by scaling from solar flares,  \cite{2017arXiv171108484T} calculate that, in the case of a stellar flare with energy 10$^{34}$ ergs, the CME/SEP effects could cause extensive ozone destruction on time-scales of years to decades. Even smaller repeated events can lead to extensive ozone destruction on century-long timescales \citep{2017ApJ...843...31Y}. However, studies of Type II radio bursts in flare stars \citep{2018ApJ...862..113C} indicate that a simple scaling from solar CME rates to CMEs in stellar flares is not consistent with their data: they conclude that this ``casts serious doubt on the assumption that a high flaring rate corresponds to a high rate of CMEs". As a result, impact of flares on ozone layer depletion remains under investigation.\\ \\
In order to assess the habitability of M dwarf planets, it is important to constrain the flare rates of M dwarfs as a function of their masses and ages. Because of this, we have been studying the flare rates of various mid and late-M dwarfs and early-L dwarfs which were observed by the $K2$ mission \citep{2014PASP..126..398H} in various campaigns (see for e.g., \citealt{2017ApJ...845...33G,2018ApJ...858...55P}). Here we present our results on superflares which we have observed on three late-M dwarfs: 2MASS J08315742+2042213 (hereafter 2M0831+2042), 2MASS J08371832+2050349 (hereafter 2M0837+2050) and 2MASS J08312608+2244586 (hereafter 2M0831+2244). In Section \ref{sec:target_characteristics}, we discuss the physical and photometric characteristics of the targets. In Section \ref{sec:data_reduction}, we present the data reduction, flare photometry and flare energy computation. We discuss the results in Section \ref{sec: discussion}.
% %
%
\section{Target Characteristics} \label{sec:target_characteristics}
The properties of our three targets are listed in Table \ref{table:properties}. We have also listed the corresponding properties of TRAPPIST-1 in the same table to enable readers to compare its properties with other stars studied in this paper. We estimated radial velocities (RVs) and $UVW$ components of space motion of our targets. The RVs were measured via cross correlation of SDSS spectra to the \cite{2007AJ....133..531B} template spectra, and the $UVW$ were calculated from photometric distances and proper motions from SDSS-2MASS-WISE coordinates, described in Schmidt et al. (2019, in prep.). \\ \\
The stars 2M0831+2042 and 2M0837+2050 have spectral types of M7 and M8 respectively \citep{2011AJ....141...97W}. Both objects are members of the open cluster Praesepe \citep{2012MNRAS.426.3419B}, which is also known as the Beehive Cluster, M44 or  NGC 2632. \cite{2018A&A...616A..10G} estimate the distance of this cluster to be 186.2 pc (distance modulus = 6.350) and its age to be $\sim$700 Myr. The M7 and M8 stars mentioned above have H$\alpha$ emission with equivalent width (hereafter EW) 9.3$\pm$0.3 and 20.3$\pm$0.6 $\AA$ respectively \citep{2015AJ....149..158S}. \\ \\
2M0831+2244 is an M9 dwarf \citep{2011AJ....141...97W} located at a distance of 74.0 pc \citep*{2018arXiv180409365G}. It has an H$\alpha$ emission with EW 7.3$\pm$0.5 $\AA$ \citep{2015AJ....149..158S}. It has a tangential velocity of $\sim$21 km s$^{-1}$ suggesting that it is younger than another late-M dwarf TRAPPIST-1 which has a tangential velocity of $\sim$60 km s$^{-1}$. Youth is also supported by its rotation period of $\sim$7 hrs measured by $K2$. The BANYAN $\Sigma$ tool \citep{2018ApJ...856...23G} suggests that it is not a member of any known nearby moving groups or stellar associations within 150 pc, by using astrometry measured by $Gaia$ and the RV we estimated. So 2M0831+2244 is probably a field star.  
% %
\begin{table*}
 \caption{Properties of targets}
 	\label{table:properties}
     \centering
      \scalebox{0.7}{
     \begin{tabular}{cccccccc}
      \hline
       \hline
        PHOTOMETRIC PROPERTIES \\
        \hline
         Target name & sp. type & $\tilde K_{p}$ & $J$ & $K$ & $i$ & H$\alpha$ EW  \\
          & & (mag) & (mag) & (mag) & (mag) & \AA \\
         \hline
         \hline
         2M0831+2042 & M7 & 19.6 & 15.56$\pm$0.06 & 14.70$\pm$0.09 & 18.51$\pm$0.01 & 9.3$\pm$0.3$^{a}$ \\
         2M0837+2050 & M8 & 20.0  & 15.90$\pm$0.07 & 14.88$\pm$0.09 & 18.80$\pm$0.01 & 20.3$\pm$0.6$^{a}$ \\
         2M0831+2244 & M9 & 19.8 & 14.91$\pm$0.04 & 13.84$\pm$0.04 & 18.77$\pm$0.01 & 7.3$\pm$0.5$^{a}$  \\
         TRAPPIST-1 & M8 & 15.9 & 11.35$\pm$0.02 & 10.30$\pm$0.02 & 15.11$\pm$0.00 & 4.9$^{b}$ \\
       \hline
      \hline
      KINEMATIC PROPERTIES \\
      \hline
       & $\mu_{\alpha}$ & $\mu_{\delta}$ & \textit{V}$_{tan}$ & RV & $U$ & $V$ & $W$ \\
        & (mas yr$^{-1}$) & (mas yr$^{-1}$) & (km s$^{-1}$) & (km s$^{-1}$) & (km s$^{-1}$) & (km s$^{-1}$) & (km s$^{-1}$) \\
      \hline
      \hline
      2M0831+2042 &-35.8$\pm$0.9 & -13.5$\pm$0.5 & 34 & 38 $\pm$11 & 41$\pm$12 & 10 $\pm$12 & 1 $\pm$12 \\
      2M0837+2050 &  -39.2$\pm$1.2 & -13.3$\pm$0.6 & 37 & 30$\pm$12 & 23$\pm$10 & 2$\pm$16 & 8.3$\pm$8.3 \\
      2M0831+2244 &  59.7$\pm$0.9 & -2.41$\pm$0.59 & 21 & 4$\pm$25 & -16 $\pm$19 & 12$\pm$8.4 & 21 $\pm$13 \\
      TRAPPIST-1 & 924$\pm$4$^{c}$ & -467$\pm$3$^{c}$ & 60 & -53 $^{d}$ & -44$^{e}$ & -67$^{e}$ & 16$^{e}$ \\
      \hline
      \hline
      EPIC IDS AND OTHER PROPERTIES \\
      \hline
      \hline
      &  EPIC ID & Member & Age & parallax & period \\
 \hline
 \hline
       2M0831+2042 & 212027121 & Praesepe & $\sim$700 Myr & 5.36$\pm$0.05 mas & 0.556$\pm$0.002 d\\
       2M0837+2050 & 212035340 & Praesepe & $\sim$700 Myr &  5.36$\pm$0.05 mas &  0.193$\pm$0.000 d \\
       2M0831+2244 & 212136544 & & & 13.5$\pm$0.6 mas & 0.292$\pm$0.001 d \\
       TRAPPIST-1 & 200164267 & & 7.6$\pm$2.2 Gyr$^{f}$  & 82.4$\pm$0.8$^{g}$ mas & 3.29 $\pm$0.07 d \\
       \hline
       \hline
       \end{tabular}}
       \\
        \textbf{Note}:\\
        i)  Spectral types are from \cite{2011AJ....141...97W} and \cite{2006PASP..118..659L}. \\
        ii) $J$ and $K$ magnitudes are from 2MASS Survey \citep{2003tmc..book.....C}. \\
        iii) $i$ magnitudes are from Pan-STARRS Survey \citep{2016arXiv161205560C}. \\
        v) The distances and proper motions are from $Gaia$ DR2 except for TRAPPIST-1. \\
        \textbf{References}:  \\
        $^{a}$\cite{2015AJ....149..158S}; $^{b}$\cite{2000AJ....120.1085G}; $^{c}$\cite{2018ApJ...862..173T}; $^{d}$\cite{2018A&A...612A..49R}; $^{e}$\cite{2009ApJ...705.1416R}; $^{f}$\cite{2017ApJ...845..110B}; $^{g}$\cite{2018ApJ...853...30V}
\end{table*}
%\label{table:properties}
%
%
\section{Data reduction and computations} \label{sec:data_reduction}
\subsection{\textit{K2} photometry}
\begin{figure} 
    \includegraphics[scale=0.5,angle=0]{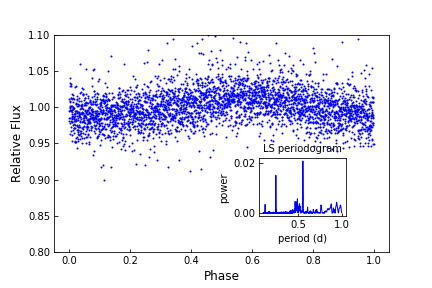} 
    \caption{The phase folded $K2$ Campaign 5 light curve of M7 dwarf 2M0831+2042 corresponding to period of 0.556 d. The LSP is also shown inside. The second peak in the LSP correponds to instrumental noise of $\sim$0.25 d.  }
    \label{fig:phase_folded_2M0831+2042}
\end{figure}

\begin{figure} 
    \includegraphics[scale=0.5,angle=0]{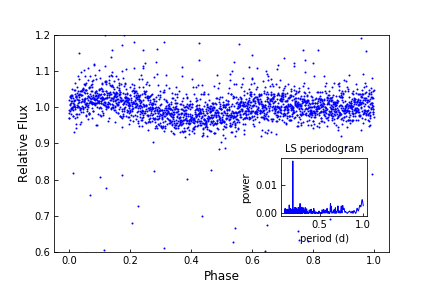} 
    \caption{The phase folded $K2$ Campaign 18 light curve of M8 dwarf 2M0837+2050 corresponding to period of 0.193 d. The LSP is also shown inside. }
    \label{fig:phase_folded_2M0837+2050}
\end{figure}

\begin{figure} 
    \includegraphics[scale=0.5,angle=0]{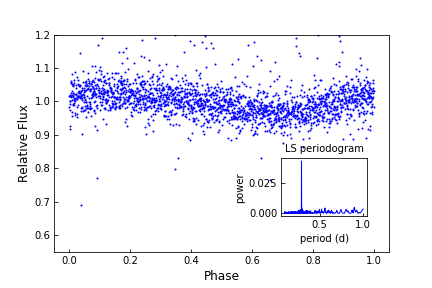} 
    \caption{The phase folded $K2$ Campaign 18 light curve of M9 dwarf 2M0831+2244 corresponding to period of 0.292 d. The LSP is also shown inside. }
    \label{fig:phase_folded_2M0831+2244}
\end{figure}

All three objects were observed by the $K2$ mission twice: once in Campaign 5 (27 April, 2015 - 10 July, 2015), and once in Campaign 18 (12 May, 2018 - 02 July, 2018). Both observations were obtained in long cadence ($\sim$30 minute) mode \citep{2010ApJ...713L.120J}. Additionally, 2M0831+2042 was observed in Campaign 16 (07 December, 2017 - 25 February, 2018). We performed point source function (psf) photometry to extract the lightcurves of our targets from their Target Pixel Files (TPFs). For this we also used the Python package `Lightkurve' \citep*{lightkurve}. The lightcurves were then detrended using the K2 Systematics Correction (`K2SC', \cite{Aigrain2016}). The median count rates of 2M0831+2042, 2M0837+2050 and 2M0831+2244 are 192, 132 and 163 counts s$^{-1}$ respectively. The \textit{Kepler} magnitude ($\tilde K_{p}$) of each object is listed in Table \ref{table:properties}. $\tilde K_{p}$ was estimated using the \cite{2015ApJ...806...30L} relation $\tilde K_{p}$ $\equiv$ 25.3 - 2.5log(count rate). We used only good quality (Q = 0) data points for flare photometry presented in this paper. Using the Lomb-Scargle periodogram (hereafter LSP), we detected periodic features with periods of 0.556$\pm$0.002, 0.193$\pm$0.000 and 0.292$\pm$0.001 d in the lightcurves of 2M0831+2042, 2M0837+2050 and 2M0831+2244 respectively. The uncertainties in the periods are based on half width at half maximum (HWHM) of the periodogram peaks \citep{2013AJ....145..148M}. These periods might be due to spot modulations of the objects: if so, the features most likely represent their rotation periods. If this is correct, the fastest rotator among our targets (2M0837+2050) rotates in a period which is shorter than 97\% of the M dwarfs which are classified as ``Class A rotators" by \cite{2016ApJ...821...93N}. Interestingly, this target turns out to be the site of the largest flare we report in this paper. The phase folded light curves of three targets are shown in Figure \ref{fig:phase_folded_2M0831+2042}, \ref{fig:phase_folded_2M0837+2050} and \ref{fig:phase_folded_2M0831+2244} respectively. The corresponding periodogram is also shown inside each figure.
\subsection{Flare detection}
\begin{figure} 
    \includegraphics[scale=0.5,angle=0]{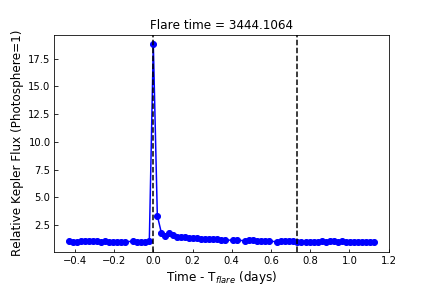} 
    \caption{The superflare observed on M7 dwarf 2M0831+2042. The blue dots represent the observed data, and the vertical dashed lines represent the start and end times of the flare. The time along the $X$-axis is centered at the peak flare time.}
    \label{fig:superflare_2M0831+2042}
\end{figure}

\begin{figure} 
    \includegraphics[scale=0.5,angle=0]{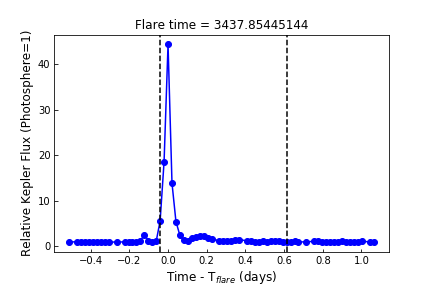} 
    \caption{The stronges superflare observed on M8 dwarf 2M0837+2050.
    The blue dots represent the observed data, and the vertical dashed lines represent the start and end times of the flare. The time along the $X$-axis is centered at the peak flare time.}
    \label{fig:superflare_2M0831+2050}
\end{figure}

\begin{figure} 
    \includegraphics[scale=0.5,angle=0]{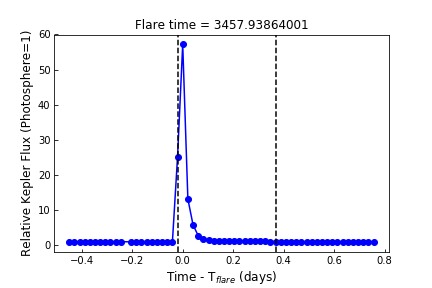} 
    \caption{The strongest superflare observed on M9 dwarf 2M0831+2244.
    The blue dots represent the observed data, and the vertical dashed lines represent the start and end times of the flare. The time along the $X$-axis is centered at the peak flare time.}
    \label{fig:superflare_2M0831+2244}
\end{figure}

\begin{figure*} 
    \includegraphics[scale=0.60,angle=0]{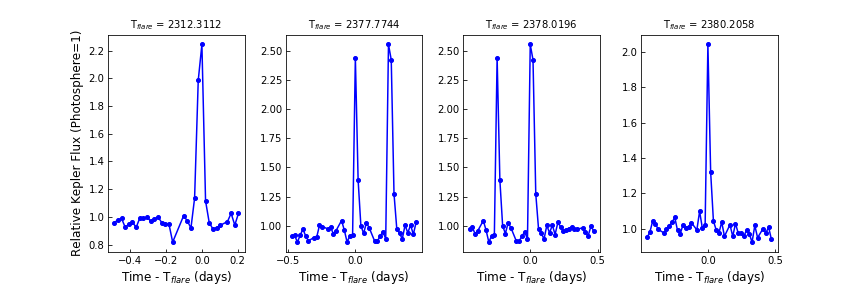} 
    \caption{Other superflares observed on 2M0837+2050 in Campaign 5 data. The blue dots represent the observed data and the time is centered at peak flare time. Though the second and the third plots are similar, the only difference is the peak flare time mentioned on the top of them, corresponding to each flare.}
    \label{fig:other_superflares_2M0837+2050}
\end{figure*}

\begin{figure*} 
    \includegraphics[scale=0.60,angle=0]{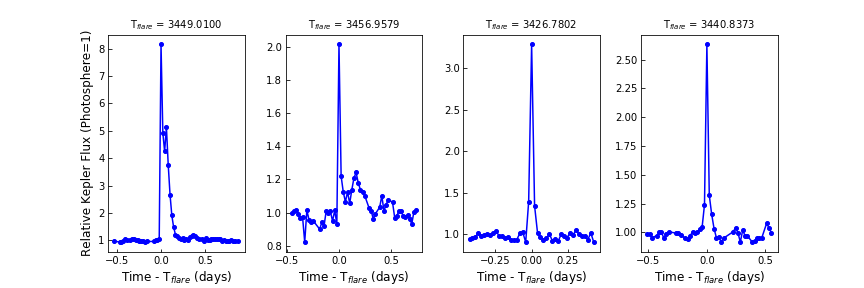} 
    \caption{Other superflares observed on 2M0831+2244. The blue dots represent the observed data and the time is centered at peak flare time. }
    \label{fig:other_superflares_2M0831+2244}
\end{figure*}

We used the method described in \cite{2012ApJ...754....4O} to identify the flares in the light curves of our targets. For each data point in the \textit{K2SC} detrended light curve, we calculated the relative flux $F_{rel,i}$, defined as:
\begin{equation}
F_{rel,i} = \frac{F_i - F_{mean}}{F_{mean}} 
\end{equation}
where \textit{F$_{i}$} is the flux in \textit{i}th epoch and F$_{mean}$ is the mean flux of the entire light curve of each target. Using the values of $F_{rel,i}$, we then calculated a statistic $\phi_{ij}$ for each consecutive observation epoch ($i,j$) as:
\begin{equation}
\phi_{ij} = \Big(\frac{F_{rel,i}}{\sigma_{i}} \Big) \times \Big(\frac{F_{rel,j}}{\sigma_{j}}\Big) , j = i+1
\end{equation}
Here \textit{$\sigma_{i}$} is the error in the flux which is associated with the \textit{i}th epoch. We then identified possible flare candidates in the light curve by using the false discovery rate analysis described in \cite{2001AJ....122.3492M}. More detailed explanation regarding this method of flare detection can be found in \cite{2012ApJ...754....4O} and \cite{2018ApJ...858...55P}. In order for the flare candidate to be qualified as a real flare event, we imposed an additional criterion, namely, that the detrended flux should exceed the photospheric level (the median count rate in the light curve) by 2.5$\sigma$.  The final flare sample was chosen by inspecting the flare light curve and pixel level data by eye. We excluded flare candidates with only a single epoch brightening. All the flares identified on our targets have at least two epoch pairs for which $\phi_{ij}$ $>$ 0 and $F_{rel,i,j}$ $>$ 0 which ensures flux brightening for multiple times. \\ \\
In the case of light curves measured in the long cadence mode, very small flares have relatively small amplitudes and last only for a few minutes (less than the duration of one long cadence at $\sim$30 minutes). They show up as single point brightening. Because of this, they do not qualify as flare candidates and hence are difficult to identify by using robust statistical techniques. \\ \\
We searched for flares on our targets in all the available light curves using our flare detection method. We identified one strong superflare on 2M0831+2042, one on  2M0837+2050, and five superflares on 2M0831+2244 in Campaign 18 light curves. We also identified four superflares on 2M0837+2050 in Campaign 5 data. The peak flare times, equivalent durations, changes in $Kepler$ magnitude ($\tilde K_{p}$) and flare durations of each superflares are listed in Table \ref{table:flare properties}. We plot the strongest superflares on our targets in Figure \ref{fig:superflare_2M0831+2042}, \ref{fig:superflare_2M0831+2050} and \ref{fig:superflare_2M0831+2244}. Likewise, we plot the remaining superflares identified on 2M0831+2050 in Figure \ref{fig:other_superflares_2M0837+2050} and those on 2M0831+2244 in Figure \ref{fig:other_superflares_2M0831+2244} respectively. In each plot, the flare flux is normalized by the median flux in the corresponding light curve. The time on the top of each plot is the peak flare (\textit{Kepler} mission) time of the corresponding flare inside the plot. 
\subsection{Flare Energies}
The most common method to estimate energy of a flare is to use its equivalent duration (ED) which is independent of the distance to the flaring object. It depends on the filter and the photospheric properties of the stars. It has units of time and is the area under the flare light curve \citep{1972Ap&SS..19...75G}. It is the time during which the flare emits the same energy as the (sub)stellar object emits when it is in its quiescent state. We followed the procedures described in \cite{2017ApJ...838...22G,2017ApJ...845...33G} to estimate the flare energies. We first estimated the photospheric spectrum of our targets by using the matching active M dwarf template \citep{2007AJ....133..531B} normalized to match Pan-STARRS $i$-band photometry \citep{2012ApJ...750...99T,2016arXiv161205560C,2016arXiv161205242M}. As a white light flare can be best described by using 10,000 K blackbody model (see for e.g., \citealt{1991ApJ...378..725H,2013ApJ...779..172G}), we modeled the flare as an 10,000 K blackbody normalized to have the same count rate through the $K2$ response curve as the photosphere of the corresponding object to estimate the energy emitted corresponding to an ED of 1 s. We emphasize that we calibrated the flare energies by extrapolating to ultra-violet and infra-red wavelengths that are not detected by $K2$. So the energies reported here are those emitted by the blackbody continuum in UV/visible/IR wavelengths. They also include atomic emission features between 430 nm and 900 nm which can be detected by $K2$ but not the emission features in UV. The blackbody continuum dominates the flare energy budget in UV/visible wavelengths \citep{1991ApJ...378..725H,2015ApJ...809...79O}. We find that a flare with ED of 1 s on 2M0831+2042 has an energy of 2.6 $\times$ 10$^{30}$ erg. Likewise, a flare with ED of 1s on 2M0837+2050 has 2.1 $\times$ 10$^{30}$ erg and a flare with ED of 1 s on 2M0831+2244 has 3.4 $\times$ 10$^{29}$ erg. We estimated the flare energies by multiplying these energies with the EDs of flares observed on corresponding targets. All the flare energies are listed in Table \ref{table:flare properties}. 
\begin{figure} 
    \includegraphics[scale=0.60,angle=0]{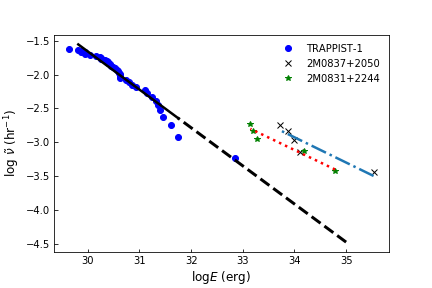} 
    \caption{Comparison of flare frequency distribution of TRAPPIST-1 with 2M0837+2050 and 2M0831+2244. The solid black line plotted over TRAPPIST-1 flare energies is fitted line using parameters from \citep{2018ApJ...858...55P} and the dashed black line represents extrapolation to energy log \textit{E} (erg) $=$ 35.}
    \label{fig:FFD_comparison}
\end{figure}
\begin{figure} 
    \includegraphics[scale=0.60,angle=0]{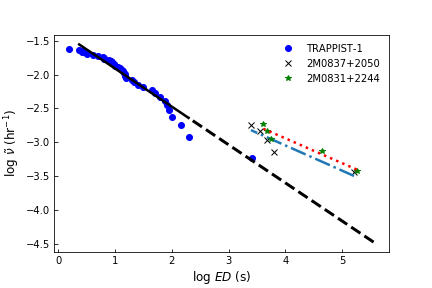} 
    \caption{Same as Figure \ref{fig:FFD_comparison}. The flare energies are replaced by corresponding EDs.}
    \label{fig:FFD_comparison_EDonly}
\end{figure}
\begin{table}
 	\caption{Properties of superflares}
 	\label{table:flare properties}
     \centering
      \scalebox{0.7}{
     \begin{tabular}{cccccc}
      \hline
       \hline
         Target & T$_{peak}$ & $\Delta \tilde K_{p}$ & ED & Energy & flare duration \\
          & (BJD - 2454833)  & & (hr) & (erg) & (d) \\
         \hline
         \hline
         2M0831+2042 & 3444.1064 &  -3.2 & 13.7 & 1.3$\times$10$^{35}$ & 0.74 \\
         2M0837+2050 & 3437.8544 & -4.1 & 46.4 & 3.5$\times$10$^{35}$ & 0.65  \\
         2M0837+2050 & 2378.0196 & -1.1 & 1.7 & 1.3$\times$10$^{34}$ & 0.06 \\
         2M0837+2050 & 2312.3112   & -1.0 & 1.3 & 1.0$\times$10$^{34}$ & 0.06 \\
         2M0837+2050 & 2377.7744  & -1.1 & 1.0 &  7.4$\times$10$^{33}$  & 0.04 \\
         2M0837+2050 & 2380.2058  & -0.9 & 0.7 &   5.2$\times$10$^{33}$ & 0.04 \\
         2M0831+2244 & 3457.9386  & -4.4 & 50.21 & 6.1 $\times$10$^{34}$  & 0.40 \\
         2M0831+2244 & 3449.0100  & -2.3 & 12.6 & 1.5$\times$10$^{34}$ & 0.50 \\
         2M0831+2244 & 3426.7801  & -1.3 & 1.6 & 1.9$\times$10$^{33}$ & 0.06 \\
         2M0831+2244 & 3456.9579   & -0.8 & 1.3 & 1.6 $\times$10$^{33}$ & 0.31  \\
         2M0831+2244 & 3440.8373  & -1.1 & 1.2 & 1.4$\times$10$^{33}$ & 0.08 \\
       \hline
      \hline
      \end{tabular}}
\end{table}
\subsection{Comparison of flare rates with TRAPPIST-1}
In Figure \ref{fig:FFD_comparison}, we compare the flare frequency distribution of TRAPPIST-1 with two of our targets: 2M0837+2050 and 2M0831+2244. This is a log-log plot of cumulative frequency($\tilde{\nu}$) of flare energies. The cumulative frequency of flares with energy $E$ is the number of flares with energies $\geq E$. The flare energies of TRAPPIST-1 are taken from \cite{2018ApJ...858...55P}. The total observation time of TRAPPIST-1 is 70.6 days and there are 39 flares with energies in the range (0.65 - 710) $\times$ 10$^{30}$ erg in the UV/visible/IR wavelengths. The total observation times of 2M0837+2050 and 2M0831+2244 are 115.58 and 112.78 days respectively, and include both Campaign 5 and 18.\\
The flare energy distribution of TRAPPIST-1 follows a power law of form 
\begin{equation} \label{eq:cu_power_law}
log\tilde{\nu} = \alpha - \beta logE
\end{equation}
with $\beta$ $\sim$0.6  \citep{2017ApJ...841..124V,2018ApJ...858...55P}. In the case of TRAPPIST-1, the black solid line represents this distribution of observed flare energies while the dashed black line represents the extrapolation of the fitted line up to flare energy of 10$^{35}$ erg to make it easier to compare with the other two targets. The extrapolation may not necessarily represent the true distribution. Likewise, the dashed lines overplotted on flare energies of 2M0837+2050 and 2M0831+2244 also represent the fitted lines. However, the lack of sufficient data point means they may not be the best representation of the real flare energy distribution of the corresponding targets.
We should note here that TRAPPIST-1 light curves were obtained in short cadence ($\sim$1 minute) mode while we do not have such light curves for 2M0837+2050 and 2M0831+2244. This is why we did not detect flares of smaller energies that are detected in the TRAPPIST-1 light curve. \\ \\
In Figure \ref{fig:FFD_comparison_EDonly}, we compare the flare rates using the EDs of flares. 
\subsection{X-ray energy emitted during the strongest flare on 2M0837+2050} \label{subsection:X-ray estimates}
To understand the relative impact of these superflares on exoplanets, we place these superflares in the solar context using the $GOES$ (Geostationary Operational Environmental Satellite) flare classification scheme.  The $GOES$ flare classification scheme (A, B, C, M, X) is based solely on the peak 1$-$8 $\AA$ soft X-ray solar flux as observed from Earth, and each letter represents an increased order of magnitude from 10$^{-8}$ W m$^{-2}$ to 10$^{-3}$ W m$^{-2}$ as observed at 1 AU. In the units of energy emitted per second, $GOES$ A1 flare corresponds to 2.8 $\times$ 10$^{22}$ erg s$^{-1}$ and $GOES$ X1 flare corresponds to  2.8 $\times$ 10$^{27}$ erg s$^{-1}$ . To estimate the flare energy in the $GOES$ bandpass, we use the energy partition from \cite{2015ApJ...809...79O} which estimates that 6\% of the bolometric flare energy is emitted in the $GOES$ 1$-$8 $\AA$ bandpass for active stars. We do not have such estimates for late-M dwarfs. To estimate the peak flare flux in the $GOES$ band, we also assume that the soft X-ray lightcurve is the same shape as the $K2$ lightcurve, and find that the peak is approximately 4 $\times$ 10$^{30}$ erg s$^{-1}$, corresponding roughly to an X10,000 class flare. 
\section{Summary and Discussion} \label{sec: discussion}
We detected strong superflares on three late-M dwarfs: 2M0831+2042 (M7 V), 2M0837+2050 (M8 V) and 2M0831+2244 (M9 V), in \textit{K2} long cadence light curves. The strong superflare observed on 2M0831+2042 has an ED of 13.7 hr and has an estimated energy of 1.3 $\times$ 10$^{35}$ erg. We detected five superflares on 2M0837+2050 with EDs in the range (0.7 - 46.4) hr and estimated energies in the range (5.2 - 350) $\times$ 10$^{33}$ erg. Likewise, we also detected five superflares on 2M0831+2244 with EDs in the range (1.2 - 50.21) hr and estimated energies in the range (1.4 - 61) $\times$ 10$^{33}$ erg. \\ \\
2M0837+2050 and 2M0831+2244 now have the highest known superflare rates among late-M dwarfs. 2M0837+2050 is $\sim$ 700 Myr old and 2M0831+2244 also appears to be young as suggested by its tangential velocity estimate. In addition, they are rapidly rotating with periods of 0.193$\pm$0.000 and 0.292$\pm$0.001 d respectively, which we detected in their $K2$ light curves using Lomb-Scargle periodogram. The high flare rates could be the result of strong magnetic dynamos enhanced by rapid rotation. Our results are in tension with those of \cite{2018arXiv180909177M} who analyzed the flare rates of a sample of 34 mid-to-late M dwarfs. They found that the flare rates are small among the slowest rotators with periods $>$70 d, but also are small among the fastest rotators with periods $<$10 d. Maximum flare rates are found among intermediate period rotators with periods of (10 - 70) d.\\
\\
The rapid rotation rates and hence high flare rates of 2M0837+2050 and 2M0831+2244 could also be the result of their presence in binary systems. \cite{2017ApJ...842...83D} measured the rotation periods of 677 low-mass stars (1 $\gtrsim$ $M$ $\gtrsim$ 0.1 $M_{\odot}$) in the Praesepe cluster by using $K2$ light curves and found that $\sim$50\% of the rapidly rotating $\gtrsim0.3 M_{\odot}$ stars are in binary systems. The sample consisted of both confirmed and candidate binary systems but there is no information regarding binarity of the remaining $\gtrsim$0.3$M_{\odot}$ fast rotators. Furthermore, \cite{2016ApJ...822...47D} found that almost all $\gtrsim$0.3 $M_{\odot}$ fast rotating stars in the Hyades cluster, with age comparable to the Praesepe cluster, are in binary systems. 
\subsection{Flare rates of $\sim$ 700 Myr TRAPPIST-1 like objects}
The flare rates of the M8 dwarf 2M0837+2050 are of particular importance for studies which are focused on the atmospheres of planets in the HZ of TRAPPIST-1. This is because both 2M0837+2050 and TRAPPIST-1 have a similar spectral type (M8) but different ages ($\sim$700 Myr, 7.6 Gyr respectively). We compare the flare frequency distributions of the two targets in Figure \ref{fig:FFD_comparison}. The energy of the largest flare on 2M0837+2050 is larger by 2.7 orders of magnitude than the largest flare observed on TRAPPIST-1 and occurs at approximately the same frequency. Our results may provide some guidance as to how large flares could have been on TRAPPIST-1 during its youth. While we do not have  enough flares to reliably constrain the flare rates, the observed flares suggest that 10$^{34}$ erg flares occur on 2M0837+2050 at a 10 times higher rate than on TRAPPIST-1. To be sure, the two M8 stars are not exactly comparable because 2M0837-2050 rotates almost 20 times faster than TRAPPIST-1. In view of this faster rotation, it is hardly surprising that the flare energy on 2M0837+2050 is hundreds of times larger than the largest flare on TRAPPIST-1. 
\\ \\
The higher superflare rate of M9 dwarf 2M0831+2244 supports the claims we made in the previous paragraph. Comparison of the tangential velocity of TRAPPIST-1 ($\sim$60 km s$^{-1}$) with that of the 2M0831+2244 ($\sim$21 km s$^{-1}$) suggests that the latter is a younger object, with an age less than 7.5 Gyr. 2M0837+2050 and 2M0831+2244 have almost identical flare frequency distributions as seen in Figure \ref{fig:FFD_comparison}. The small differences between them might be due to differences in the rotation rates and/or the effective temperatures. The larger EW of H$\alpha$ emission of 2M0837+2050 is also consistent with a higher flare rate than 2M0831+2244. \\ \\
In Figure \ref{fig:FFD_comparison_EDonly}, we compare the rates using only the EDs of flares. It suggests that the rates are almost similar for 2M0837+2050 and 2M0831+2244.  However, we have only one data point of TRAPPIST-1 for which the ED value overlaps with the EDs of 2M0837+2050 and 2M0831+2244 flares. Though this single data point suggests that flares with comparable EDs occur more frequently on younger targets, we cannot perform any further statistical analysis and conclude anything at this point.   
\subsection{Particle flux associated with the largest superflare on 2M0837+2050}
In Section \ref{subsection:X-ray estimates}, we estimated that the largest superflare from 2M0837+2050 is approximately X10,000 class in the GOES classification scheme of solar flares. For comparison, the Carrington event of 1859, probably the largest solar flare ever recorded, was $\sim$X45 class\citep{2013JSWSC...3A..31C} . \cite{2010AsBio..10..751S} estimate that the Great AD Leo flare \citep{1991ApJ...378..725H} was X2300 class, and the extreme flare from young M dwarf binary DG CVn that triggered the Swift Burst Alert Telescope was estimated to be X600,000 class \citep{2016ApJ...832..174O}. Solar flares $>$X10 class have a $\sim$100\% probability of accompaniment by a CME \citep{2006ApJ...650L.143Y}, but events larger than the Carrington event have never been observed from the Sun and the relationship between CME properties and superflares is not yet known (e.g., \cite{2011SoPh..268..195A}). Extending solar scaling relations \citep{2012ApJ...756L..29C} out to this observationally unconstrained superflare regime, we find that the $>$10 MeV proton flux expected at 0.02 AU (approximately the location of Trappist-1 d) is 4.9 $\times$ 10$^{11}$ pfu (proton cm$^{-2}$ s$^{-1}$ sr$^{-1}$). For comparison, the $>$10 MeV proton flux that \cite{2010AsBio..10..751S} estimated would have impacted a hypothetical HZ planet at 0.16 AU for the Great AD Leo flare was 6 $\times$ 10$^{8}$ pfu, and this proton flux was sufficient to destroy 94\% of the O$_{3}$ column density of a modeled Earth-like planet. 
\\ \\
\subsection{Comparison of superflare on 2M0837+2050 with that on M2 dwarf GSC 8056-0482 for possible UV flux estimates}
It is also instructive to compare the flare we discovered on one of our targets with a flare on another M dwarf, although they differ in their global properties. The superflare which we have discovered on the M8 dwarf 2M0837+2050 is larger by $\sim$100$\times$ than the $\sim$4 $\times$ 10$^{33}$ erg flare observed on GSC 8056-0482 by \cite{2018ApJ...867...70L}. The latter star (hereafter GSC) differs from our target star 2M0837+2050 in that it is warmer (sp. type = M2, $T_{eff}$ $\sim$ 3440 K; \citealt{2013ApJS..208....9P}) and younger ($\sim$40 Myr; \citealt{2014AJ....147..146K}). In view of these differences, we need to exercise caution in making comparisons. The FUV energy of the superflare on GSC was estimated to be 10$^{32.1}$ erg and is the largest energy flare ever observed in the FUV with the \textit{Hubble Space Telescope (HST)}. Assuming the same ratio (18\%: \citealt{2018ApJ...867...70L}) of FUV energy to bolometric energy of the GSC flare, we estimate that the FUV energy of our superflare on 2M0837+2050 is $>$6.3$\times$ 10$^{34}$ erg (we do not have the estimate of total bolometric energy of the flare on 2M0837+2050). This suggests that the FUV energy emitted during the superflare on 2M0837+2050 is greater than that on GSC by more than 2.7 orders of magnitude. \\ \\ 
\cite{2018ApJ...867...70L} analyzed FUV flares on 12 M dwarfs in the $\sim$40 Myr Tuc-Hor young moving group, with spectral types M0.0-M2.3. They found that the flares on those M dwarfs were 100-1000$\times$ more energetic than those on field age ($\sim$1 - 9 Gyr) M dwarfs. Combining their results with those we obtained in this paper, it is possible that strong flares with FUV energy in excess of 10$^{36}$ erg can occur on the young $\sim$40 Myr M8 dwarfs. If superflares with energies of 10$^{36}$ ergs occur frequently on young M8 dwarfs, the large amount of UV  flux in the superflares may have serious consequences on the atmospheres of planets in HZ including complete loss of O$_{3}$ column \citep{2010AsBio..10..751S,2017arXiv171108484T,2017ApJ...843...31Y,2018ApJ...860L..30H}. Such results will be very important to study the environment arount planet-hosting stars like TRAPPIST-1 when they were very young.
\subsection{Timescales of planet formation and life on Earth}
The median lifetime of circumstellar disks, the planets forming regions, is $\sim$3 Myr and the dissipiation rate is slower in case of low mass stars than that of high-mass stars \citep{2011ARA&A..49...67W}. The disk lifetimes of M dwarfs are of the order of $\sim$10 Myr. This is inferred from the fact that the dust disk around young M dwarfs are similar to those around T-Tauri stars whose disk life timescales are $\sim$10 Myr \citep{2006ApJ...643..501B,2011ASPC..448..469P}. On the other hand, the timescale of formation of rocky planets in our solar system is $\sim$10-120 Myr and that of gas giants is $<$5 Myr. A longer disk lifetime of M dwarfs may indicate a different timescale of planet formation around them.  \citep{2013AN....334...57A}. It is very likely the M dwarf planets might have already formed at ages of the Praesepe cluster.  \\ \\
On the other hand, the emergence of life also requires a minimum amount of time: in the only system for which we have data, it appears that an interval of some 200 Myr elapsed between the time of Earth formation and the time when the oldest known life emerged \citep{Dodd:2017aa}. If this time-scale is relevant to emergence of life on exoplanets, then the level of flare activity on a star whose age is less than 200 Myr may not be relevant to astrobiology.
The timescale for decay in superflare rate may also set timescale for life emergence. The latter also depends on planet properties such as the presence of magnetic fields.
%
\begin{comment}
Although it is a matter of physical interest to understand how a flare might negatively impact a planetary atmosphere on time-scales of tens of Myr, the conclusion from such studies are not necessarily relevant to astrobiology. The reason is that the emergence of life also requires a minimum amount of time: in the only system for which we have data, it appears that an interval of some 200 Myr elapsed between the time of Earth formation and the time when the oldest known life emerged \citep{Dodd:2017aa}. If this time-scale is relevant to emergence of life on exoplanets, then the level of flare activity on a star whose age is less than 200 Myr may not be relevant to astrobiology.
\end{comment}
%
%
\subsection{How can the superflares be beneficial to studies regarding CMEs associated with stellar flares?}
The presence of strong superflares highlights the importance of studying the distribution of magnetic field strengths on late-M dwarfs. Information about field strengths is necessary if we are to determine whether they are strong enough to supress the CMEs associated with the strong superflares. For example, \cite{2018ApJ...862...93A} estimated that a large-scale dipolar field of strength 75 G is sufficient to supress the escape of the largest solar-like CMEs with kinetic energies of $\sim$3 $\times$10$^{32}$ erg.  Assuming that the requisite field strength $B$ to suppress a CME of a certain kinetic energy (KE) scales in such a way that $B^{2}$/8$\pi$ $\sim$ KE, \cite{2018arXiv181104149M} have argued that a global stellar field of 750 G could suffice to suppress CMEs in Trappist-1 with KE = 3$\times$10$^{34}$ ergs. Using the same scaling, magnetic fields of order 7.5 kG may be required to suppress CMEs with KE = 3$\times$10$^{36}$ ergs. Such supression of energetic CMEs seems possible in some M dwarfs similar to WX UMa on which strong mean magnetic fields of strength 7.0 kG exist \citep{2017NatAs...1E.184S}. But we do not know the upper limit of the flare energy and the KE of CMEs that can be emitted by an M dwarf of given mass and age. Hence it is important to the study of evolution of M dwarf flares through time to know the maximum flare energies that can be produced by each spectral type at different ages.  In addition, we need more simultaneous observations of M dwarf flares at multiple wavelengths to constrain the flare energy distribution, which together with the maximum flare energies will enable us to estimate the total energy budgets received by the planets from their parent stars. Such energies will be valuable inputs to exoplanet atmosphere and climate models. If information can be obtained as to the time $T$ which must elapse before life emerges on an exoplanet, then the results of atmosphere and climate models at times $>T$ could help put constraints on the habitability of planets in the HZ of M dwarfs.
\subsection{Possible benefits of superflares to the planets in the HZ of M dwarfs}
The superflares could also be beneficial to the planets in the HZ of M dwarfs. In an Earth-like planet orbiting within the classical HZ of a solar-type star or M dwarf, an H/He envelope having mass-fraction (ratio of mass of envelope to mass of core, $M_{env}/M_{core}$) of the order of 1\%  may lead to very high surface temperatures and pressures unsuitable for the existence of liquid water. However, liquid water can be retained if the H/He envelope mass-fraction can be reduced to $\ll$10$^{-3}$ via photoevaporation or other mechanisms (\citealt{2016MNRAS.459.4088O} and references therein).   In this regard, the superflares may be helpful to strip off the thick H/He envelope if it is present. In addition, the superflares could help in producing haze forming monomers through photolysis of methane, in planets whose atmosphere is dominated by methane. The hazes might shield the planet's surface from UV radiation which in certain circumstances can be harmful to life to life \citep{2017arXiv171108484T}. Moreover, enhancement of photon fluxes is not always harmful to life. Certain UV photons can be beneficial for the onset of life by contributing to generation of the bases which occur in nucleic acids \citep{2016NatGe...9..452A,2017ApJ...843..110R}. And as another example of positive consequences of enhanced photon fluxes, we note that the optical photons which are enhanced during flares will increase the effectiveness of oxygenic photosynthesis in a planet lying in the HZ of a flare star \citep{2018ApJ...865..101M}.\\

\begin{comment}
In addition, the mass-period relation analyzed by \cite{2016ApJ...821...93N} also suggests that the stars in 2M1027+0629 binary system are likely to have masses below the full convection limit (0.35 $M_{\odot}$; \citealt{1997A&A...327.1039C}). 
\end{comment}
% %
\section*{Acknowledgements}
The material in this paper is based upon work supported by NASA under award Nos. NNX15AV64G, NNX16AE55G and NNX16AJ22G. Some/all of the data presented in this paper were obtained from the Mikulski Archive for Space Telescopes (MAST). STScI is operated by the Association of Universities for Research in Astronomy, Inc., under NASA contract NAS5-26555. This paper includes data collected by the Kepler mission. Funding for the Kepler mission is provided by the NASA Science Mission directorate. This work has made use of data from the European Space Agency (ESA) mission
{\it Gaia} (\url{https://www.cosmos.esa.int/gaia}), processed by the {\it Gaia}
Data Processing and Analysis Consortium (DPAC,
\url{https://www.cosmos.esa.int/web/gaia/dpac/consortium}). Funding for the DPAC
has been provided by national institutions, in particular the institutions
participating in the {\it Gaia} Multilateral Agreement. This research has made use of the VizieR catalogue access tool, CDS,
 Strasbourg, France. The original description of the VizieR service was
 published in A\&AS 143, 23. This work made use of the \url{http://gaia-kepler.fun}  crossmatch database created by Megan Bedell. \\

$Softwares$: Python, IPython \cite{2007CSE.....9c..21P}, Astropy \citep{2013A&A...558A..33A}, Matplotlib \citep{Hunter:2007}, Numpy \citep{Oliphant:2015:GN:2886196}, Lightkurve \citep*{lightkurve}, K2SC \citep{Aigrain2016}, Jupyter \citep{Kluyver:2016aa}
\bibliographystyle{mnras}
\bibliography{/Users/rishipaudel/GoogleDrive/Research/astrobib}

\begin{thebibliography}{}
\makeatletter
\relax
\def\mn@urlcharsother{\let\do\@makeother \do\$\do\&\do\#\do\^\do\_\do\%\do\~}
\def\mn@doi{\begingroup\mn@urlcharsother \@ifnextchar [ {\mn@doi@}
  {\mn@doi@[]}}
\def\mn@doi@[#1]#2{\def\@tempa{#1}\ifx\@tempa\@empty \href
  {http://dx.doi.org/#2} {doi:#2}\else \href {http://dx.doi.org/#2} {#1}\fi
  \endgroup}
\def\mn@eprint#1#2{\mn@eprint@#1:#2::\@nil}
\def\mn@eprint@arXiv#1{\href {http://arxiv.org/abs/#1} {{\tt arXiv:#1}}}
\def\mn@eprint@dblp#1{\href {http://dblp.uni-trier.de/rec/bibtex/#1.xml}
  {dblp:#1}}
\def\mn@eprint@#1:#2:#3:#4\@nil{\def\@tempa {#1}\def\@tempb {#2}\def\@tempc
  {#3}\ifx \@tempc \@empty \let \@tempc \@tempb \let \@tempb \@tempa \fi \ifx
  \@tempb \@empty \def\@tempb {arXiv}\fi \@ifundefined
  {mn@eprint@\@tempb}{\@tempb:\@tempc}{\expandafter \expandafter \csname
  mn@eprint@\@tempb\endcsname \expandafter{\@tempc}}}

\bibitem[\protect\citeauthoryear{{Aarnio}, {Stassun}, {Hughes}  \&
  {McGregor}}{{Aarnio} et~al.}{2011}]{2011SoPh..268..195A}
{Aarnio} A.~N.,  {Stassun} K.~G.,  {Hughes} W.~J.,   {McGregor} S.~L.,  2011,
  \mn@doi [\solphys] {10.1007/s11207-010-9672-7}, \href
  {http://adsabs.harvard.edu/abs/2011SoPh..268..195A} {268, 195}

\bibitem[\protect\citeauthoryear{Aigrain, Parviainen  \& Pope}{Aigrain
  et~al.}{2016}]{Aigrain2016}
Aigrain S.,  Parviainen H.,   Pope B.,  2016

\bibitem[\protect\citeauthoryear{{Airapetian}, {Glocer}, {Gronoff},
  {H{\'e}brard}  \& {Danchi}}{{Airapetian} et~al.}{2016}]{2016NatGe...9..452A}
{Airapetian} V.~S.,  {Glocer} A.,  {Gronoff} G.,  {H{\'e}brard} E.,   {Danchi}
  W.,  2016, \mn@doi [Nature Geoscience] {10.1038/ngeo2719}, \href
  {http://adsabs.harvard.edu/abs/2016NatGe...9..452A} {9, 452}

\bibitem[\protect\citeauthoryear{{Alvarado-G{\'o}mez}, {Drake}, {Cohen},
  {Moschou}  \& {Garraffo}}{{Alvarado-G{\'o}mez}
  et~al.}{2018}]{2018ApJ...862...93A}
{Alvarado-G{\'o}mez} J.~D.,  {Drake} J.~J.,  {Cohen} O.,  {Moschou} S.~P.,
  {Garraffo} C.,  2018, \mn@doi [\apj] {10.3847/1538-4357/aacb7f}, \href
  {http://adsabs.harvard.edu/abs/2018ApJ...862...93A} {862, 93}

\bibitem[\protect\citeauthoryear{{Apai}}{{Apai}}{2013}]{2013AN....334...57A}
{Apai} D.,  2013, \mn@doi [Astronomische Nachrichten] {10.1002/asna.201211780},
  \href {http://adsabs.harvard.edu/abs/2013AN....334...57A} {334, 57}

\bibitem[\protect\citeauthoryear{{Astropy Collaboration} et~al.,}{{Astropy
  Collaboration} et~al.}{2013}]{2013A&A...558A..33A}
{Astropy Collaboration} et~al., 2013, \mn@doi [\aap]
  {10.1051/0004-6361/201322068}, \href
  {http://adsabs.harvard.edu/abs/2013A%26A...558A..33A} {558, A33}

\bibitem[\protect\citeauthoryear{{Barclay}, {Pepper}  \& {Quintana}}{{Barclay}
  et~al.}{2018}]{2018ApJS..239....2B}
{Barclay} T.,  {Pepper} J.,   {Quintana} E.~V.,  2018, \mn@doi [\apjs]
  {10.3847/1538-4365/aae3e9}, \href
  {http://adsabs.harvard.edu/abs/2018ApJS..239....2B} {239, 2}

\bibitem[\protect\citeauthoryear{{Bochanski}, {West}, {Hawley}  \&
  {Covey}}{{Bochanski} et~al.}{2007}]{2007AJ....133..531B}
{Bochanski} J.~J.,  {West} A.~A.,  {Hawley} S.~L.,   {Covey} K.~R.,  2007,
  \mn@doi [\aj] {10.1086/510240}, \href
  {http://adsabs.harvard.edu/abs/2007AJ....133..531B} {133, 531}

\bibitem[\protect\citeauthoryear{{Boss}}{{Boss}}{2006}]{2006ApJ...643..501B}
{Boss} A.~P.,  2006, \mn@doi [\apj] {10.1086/501522}, \href
  {http://adsabs.harvard.edu/abs/2006ApJ...643..501B} {643, 501}

\bibitem[\protect\citeauthoryear{{Boudreault}, {Lodieu}, {Deacon}  \&
  {Hambly}}{{Boudreault} et~al.}{2012}]{2012MNRAS.426.3419B}
{Boudreault} S.,  {Lodieu} N.,  {Deacon} N.~R.,   {Hambly} N.~C.,  2012,
  \mn@doi [\mnras] {10.1111/j.1365-2966.2012.21854.x}, \href
  {http://adsabs.harvard.edu/abs/2012MNRAS.426.3419B} {426, 3419}

\bibitem[\protect\citeauthoryear{{Burgasser} \& {Mamajek}}{{Burgasser} \&
  {Mamajek}}{2017}]{2017ApJ...845..110B}
{Burgasser} A.~J.,  {Mamajek} E.~E.,  2017, \mn@doi [\apj]
  {10.3847/1538-4357/aa7fea}, \href
  {http://adsabs.harvard.edu/abs/2017ApJ...845..110B} {845, 110}

\bibitem[\protect\citeauthoryear{{Chambers} et~al.,}{{Chambers}
  et~al.}{2016}]{2016arXiv161205560C}
{Chambers} K.~C.,  et~al., 2016, preprint, \href
  {http://adsabs.harvard.edu/abs/2016arXiv161205560C} {} (\mn@eprint {arXiv}
  {1612.05560})

\bibitem[\protect\citeauthoryear{{Clements}, {Henry}, {Hosey}, {Jao},
  {Silverstein}, {Winters}, {Dieterich}  \& {Riedel}}{{Clements}
  et~al.}{2017}]{2017AJ....154..124C}
{Clements} T.~D.,  {Henry} T.~J.,  {Hosey} A.~D.,  {Jao} W.-C.,  {Silverstein}
  M.~L.,  {Winters} J.~G.,  {Dieterich} S.~B.,   {Riedel} A.~R.,  2017, \mn@doi
  [\aj] {10.3847/1538-3881/aa8464}, \href
  {http://adsabs.harvard.edu/abs/2017AJ....154..124C} {154, 124}

\bibitem[\protect\citeauthoryear{{Cliver} \& {Dietrich}}{{Cliver} \&
  {Dietrich}}{2013}]{2013JSWSC...3A..31C}
{Cliver} E.~W.,  {Dietrich} W.~F.,  2013, \mn@doi [Journal of Space Weather and
  Space Climate] {10.1051/swsc/2013053}, \href
  {http://adsabs.harvard.edu/abs/2013JSWSC...3A..31C} {3, A31}

\bibitem[\protect\citeauthoryear{{Cliver}, {Ling}, {Belov}  \&
  {Yashiro}}{{Cliver} et~al.}{2012}]{2012ApJ...756L..29C}
{Cliver} E.~W.,  {Ling} A.~G.,  {Belov} A.,   {Yashiro} S.,  2012, \mn@doi
  [\apjl] {10.1088/2041-8205/756/2/L29}, \href
  {http://adsabs.harvard.edu/abs/2012ApJ...756L..29C} {756, L29}

\bibitem[\protect\citeauthoryear{{Crosley} \& {Osten}}{{Crosley} \&
  {Osten}}{2018}]{2018ApJ...862..113C}
{Crosley} M.~K.,  {Osten} R.~A.,  2018, \mn@doi [\apj]
  {10.3847/1538-4357/aacf02}, \href
  {http://adsabs.harvard.edu/abs/2018ApJ...862..113C} {862, 113}

\bibitem[\protect\citeauthoryear{{Cutri} et~al.,}{{Cutri}
  et~al.}{2003}]{2003tmc..book.....C}
{Cutri} R.~M.,  et~al., 2003, {2MASS All Sky Catalog of point sources.}

\bibitem[\protect\citeauthoryear{{Davenport}}{{Davenport}}{2016}]{2016ApJ...829...23D}
{Davenport} J.~R.~A.,  2016, \mn@doi [\apj] {10.3847/0004-637X/829/1/23}, \href
  {http://adsabs.harvard.edu/abs/2016ApJ...829...23D} {829, 23}

\bibitem[\protect\citeauthoryear{Dodd, Papineau, Grenne, Slack, Rittner,
  Pirajno, O'Neil  \& Little}{Dodd et~al.}{2017}]{Dodd:2017aa}
Dodd M.~S.,  Papineau D.,  Grenne T.,  Slack J.~F.,  Rittner M.,  Pirajno F.,
  O'Neil J.,   Little C. T.~S.,  2017, Nature, 543, 60 EP

\bibitem[\protect\citeauthoryear{{Douglas}, {Ag{\"u}eros}, {Covey}, {Cargile},
  {Barclay}, {Cody}, {Howell}  \& {Kopytova}}{{Douglas}
  et~al.}{2016}]{2016ApJ...822...47D}
{Douglas} S.~T.,  {Ag{\"u}eros} M.~A.,  {Covey} K.~R.,  {Cargile} P.~A.,
  {Barclay} T.,  {Cody} A.,  {Howell} S.~B.,   {Kopytova} T.,  2016, \mn@doi
  [\apj] {10.3847/0004-637X/822/1/47}, \href
  {http://adsabs.harvard.edu/abs/2016ApJ...822...47D} {822, 47}

\bibitem[\protect\citeauthoryear{{Douglas}, {Ag{\"u}eros}, {Covey}  \&
  {Kraus}}{{Douglas} et~al.}{2017}]{2017ApJ...842...83D}
{Douglas} S.~T.,  {Ag{\"u}eros} M.~A.,  {Covey} K.~R.,   {Kraus} A.,  2017,
  \mn@doi [\apj] {10.3847/1538-4357/aa6e52}, \href
  {http://adsabs.harvard.edu/abs/2017ApJ...842...83D} {842, 83}

\bibitem[\protect\citeauthoryear{{Dressing} \& {Charbonneau}}{{Dressing} \&
  {Charbonneau}}{2015}]{2015ApJ...807...45D}
{Dressing} C.~D.,  {Charbonneau} D.,  2015, \mn@doi [\apj]
  {10.1088/0004-637X/807/1/45}, \href
  {http://adsabs.harvard.edu/abs/2015ApJ...807...45D} {807, 45}

\bibitem[\protect\citeauthoryear{{Gagn{\'e}} et~al.,}{{Gagn{\'e}}
  et~al.}{2018}]{2018ApJ...856...23G}
{Gagn{\'e}} J.,  et~al., 2018, \mn@doi [\apj] {10.3847/1538-4357/aaae09}, \href
  {http://adsabs.harvard.edu/abs/2018ApJ...856...23G} {856, 23}

\bibitem[\protect\citeauthoryear{{Gaia Collaboration}, {Brown}, {Vallenari},
  {Prusti}, {de Bruijne}, {Babusiaux}  \& {Bailer-Jones}}{{Gaia Collaboration}
  et~al.}{2018a}]{2018arXiv180409365G}
{Gaia Collaboration} {Brown} A.~G.~A.,  {Vallenari} A.,  {Prusti} T.,  {de
  Bruijne} J.~H.~J.,  {Babusiaux} C.,   {Bailer-Jones} C.~A.~L.,  2018a,
  preprint, \href {http://adsabs.harvard.edu/abs/2018arXiv180409365G} {}
  (\mn@eprint {arXiv} {1804.09365})

\bibitem[\protect\citeauthoryear{{Gaia Collaboration} et~al.,}{{Gaia
  Collaboration} et~al.}{2018b}]{2018A&A...616A..10G}
{Gaia Collaboration} et~al., 2018b, \mn@doi [\aap]
  {10.1051/0004-6361/201832843}, \href
  {http://adsabs.harvard.edu/abs/2018A%26A...616A..10G} {616, A10}

\bibitem[\protect\citeauthoryear{{Gershberg}}{{Gershberg}}{1972}]{1972Ap&SS..19...75G}
{Gershberg} R.~E.,  1972, \mn@doi [\apss] {10.1007/BF00643168}, \href
  {http://adsabs.harvard.edu/abs/1972Ap%26SS..19...75G} {19, 75}

\bibitem[\protect\citeauthoryear{{Gillon} et~al.,}{{Gillon}
  et~al.}{2016}]{2016Natur.533..221G}
{Gillon} M.,  et~al., 2016, \mn@doi [\nat] {10.1038/nature17448}, \href
  {http://adsabs.harvard.edu/abs/2016Natur.533..221G} {533, 221}

\bibitem[\protect\citeauthoryear{{Gillon} et~al.,}{{Gillon}
  et~al.}{2017}]{2017Natur.542..456G}
{Gillon} M.,  et~al., 2017, \mn@doi [\nat] {10.1038/nature21360}, \href
  {http://adsabs.harvard.edu/abs/2017Natur.542..456G} {542, 456}

\bibitem[\protect\citeauthoryear{{Gizis}, {Monet}, {Reid}, {Kirkpatrick},
  {Liebert}  \& {Williams}}{{Gizis} et~al.}{2000}]{2000AJ....120.1085G}
{Gizis} J.~E.,  {Monet} D.~G.,  {Reid} I.~N.,  {Kirkpatrick} J.~D.,  {Liebert}
  J.,   {Williams} R.~J.,  2000, \mn@doi [\aj] {10.1086/301456}, \href
  {http://adsabs.harvard.edu/abs/2000AJ....120.1085G} {120, 1085}

\bibitem[\protect\citeauthoryear{{Gizis}, {Burgasser}, {Berger}, {Williams},
  {Vrba}, {Cruz}  \& {Metchev}}{{Gizis} et~al.}{2013}]{2013ApJ...779..172G}
{Gizis} J.~E.,  {Burgasser} A.~J.,  {Berger} E.,  {Williams} P.~K.~G.,  {Vrba}
  F.~J.,  {Cruz} K.~L.,   {Metchev} S.,  2013, \mn@doi [\apj]
  {10.1088/0004-637X/779/2/172}, \href
  {http://adsabs.harvard.edu/abs/2013ApJ...779..172G} {779, 172}

\bibitem[\protect\citeauthoryear{{Gizis}, {Paudel}, {Schmidt}, {Williams}  \&
  {Burgasser}}{{Gizis} et~al.}{2017a}]{2017ApJ...838...22G}
{Gizis} J.~E.,  {Paudel} R.~R.,  {Schmidt} S.~J.,  {Williams} P.~K.~G.,
  {Burgasser} A.~J.,  2017a, \mn@doi [\apj] {10.3847/1538-4357/aa6197}, \href
  {http://adsabs.harvard.edu/abs/2017ApJ...838...22G} {838, 22}

\bibitem[\protect\citeauthoryear{{Gizis}, {Paudel}, {Mullan}, {Schmidt},
  {Burgasser}  \& {Williams}}{{Gizis} et~al.}{2017b}]{2017ApJ...845...33G}
{Gizis} J.~E.,  {Paudel} R.~R.,  {Mullan} D.,  {Schmidt} S.~J.,  {Burgasser}
  A.~J.,   {Williams} P.~K.~G.,  2017b, \mn@doi [\apj]
  {10.3847/1538-4357/aa7da0}, \href
  {http://adsabs.harvard.edu/abs/2017ApJ...845...33G} {845, 33}

\bibitem[\protect\citeauthoryear{{Hawley} \& {Pettersen}}{{Hawley} \&
  {Pettersen}}{1991}]{1991ApJ...378..725H}
{Hawley} S.~L.,  {Pettersen} B.~R.,  1991, \mn@doi [\apj] {10.1086/170474},
  \href {http://adsabs.harvard.edu/abs/1991ApJ...378..725H} {378, 725}

\bibitem[\protect\citeauthoryear{{Howard} et~al.,}{{Howard}
  et~al.}{2018}]{2018ApJ...860L..30H}
{Howard} W.~S.,  et~al., 2018, \mn@doi [\apjl] {10.3847/2041-8213/aacaf3},
  \href {http://adsabs.harvard.edu/abs/2018ApJ...860L..30H} {860, L30}

\bibitem[\protect\citeauthoryear{{Howell} et~al.,}{{Howell}
  et~al.}{2014}]{2014PASP..126..398H}
{Howell} S.~B.,  et~al., 2014, \mn@doi [\pasp] {10.1086/676406}, \href
  {http://adsabs.harvard.edu/abs/2014PASP..126..398H} {126, 398}

\bibitem[\protect\citeauthoryear{Hunter}{Hunter}{2007}]{Hunter:2007}
Hunter J.~D.,  2007, \mn@doi [Computing In Science \& Engineering]
  {10.1109/MCSE.2007.55}, 9, 90

\bibitem[\protect\citeauthoryear{{Jenkins} et~al.,}{{Jenkins}
  et~al.}{2010}]{2010ApJ...713L.120J}
{Jenkins} J.~M.,  et~al., 2010, \mn@doi [\apjl] {10.1088/2041-8205/713/2/L120},
  \href {http://adsabs.harvard.edu/abs/2010ApJ...713L.120J} {713, L120}

\bibitem[\protect\citeauthoryear{Kluyver et~al.,}{Kluyver
  et~al.}{2016}]{Kluyver:2016aa}
Kluyver T.,  et~al., 2016, in Loizides F.,  Schmidt B.,  eds, Positioning and
  Power in Academic Publishing: Players, Agents and Agendas. pp 87 -- 90

\bibitem[\protect\citeauthoryear{{Kraus}, {Shkolnik}, {Allers}  \&
  {Liu}}{{Kraus} et~al.}{2014}]{2014AJ....147..146K}
{Kraus} A.~L.,  {Shkolnik} E.~L.,  {Allers} K.~N.,   {Liu} M.~C.,  2014,
  \mn@doi [\aj] {10.1088/0004-6256/147/6/146}, \href
  {http://adsabs.harvard.edu/abs/2014AJ....147..146K} {147, 146}

\bibitem[\protect\citeauthoryear{{Lacy}, {Moffett}  \& {Evans}}{{Lacy}
  et~al.}{1976}]{1976ApJS...30...85L}
{Lacy} C.~H.,  {Moffett} T.~J.,   {Evans} D.~S.,  1976, \mn@doi [\apjs]
  {10.1086/190358}, \href {http://adsabs.harvard.edu/abs/1976ApJS...30...85L}
  {30, 85}

\bibitem[\protect\citeauthoryear{{Liebert} \& {Gizis}}{{Liebert} \&
  {Gizis}}{2006}]{2006PASP..118..659L}
{Liebert} J.,  {Gizis} J.~E.,  2006, \mn@doi [\pasp] {10.1086/503333}, \href
  {http://adsabs.harvard.edu/abs/2006PASP..118..659L} {118, 659}

\bibitem[\protect\citeauthoryear{{Loyd}, {Shkolnik}, {Schneider}, {Barman},
  {Meadows}, {Pagano}  \& {Peacock}}{{Loyd} et~al.}{2018}]{2018ApJ...867...70L}
{Loyd} R.~O.~P.,  {Shkolnik} E.~L.,  {Schneider} A.~C.,  {Barman} T.~S.,
  {Meadows} V.~S.,  {Pagano} I.,   {Peacock} S.,  2018, \mn@doi [\apj]
  {10.3847/1538-4357/aae2ae}, \href
  {http://adsabs.harvard.edu/abs/2018ApJ...867...70L} {867, 70}

\bibitem[\protect\citeauthoryear{{Luger} et~al.,}{{Luger}
  et~al.}{2017}]{2017NatAs...1E.129L}
{Luger} R.,  et~al., 2017, \mn@doi [Nature Astronomy]
  {10.1038/s41550-017-0129}, \href
  {http://adsabs.harvard.edu/abs/2017NatAs...1E.129L} {1, 0129}

\bibitem[\protect\citeauthoryear{{Lund}, {Handberg}, {Davies}, {Chaplin}  \&
  {Jones}}{{Lund} et~al.}{2015}]{2015ApJ...806...30L}
{Lund} M.~N.,  {Handberg} R.,  {Davies} G.~R.,  {Chaplin} W.~J.,   {Jones}
  C.~D.,  2015, \mn@doi [\apj] {10.1088/0004-637X/806/1/30}, \href
  {http://adsabs.harvard.edu/abs/2015ApJ...806...30L} {806, 30}

\bibitem[\protect\citeauthoryear{{Magnier} et~al.,}{{Magnier}
  et~al.}{2016}]{2016arXiv161205242M}
{Magnier} E.~A.,  et~al., 2016, preprint, \href
  {http://adsabs.harvard.edu/abs/2016arXiv161205242M} {} (\mn@eprint {arXiv}
  {1612.05242})

\bibitem[\protect\citeauthoryear{{Mighell} \& {Plavchan}}{{Mighell} \&
  {Plavchan}}{2013}]{2013AJ....145..148M}
{Mighell} K.~J.,  {Plavchan} P.,  2013, \mn@doi [\aj]
  {10.1088/0004-6256/145/6/148}, \href
  {http://adsabs.harvard.edu/abs/2013AJ....145..148M} {145, 148}

\bibitem[\protect\citeauthoryear{{Miller} et~al.,}{{Miller}
  et~al.}{2001}]{2001AJ....122.3492M}
{Miller} C.~J.,  et~al., 2001, \mn@doi [\aj] {10.1086/324109}, \href
  {http://adsabs.harvard.edu/abs/2001AJ....122.3492M} {122, 3492}

\bibitem[\protect\citeauthoryear{{Mondrik}, {Newton}  \& {Irwin}}{{Mondrik}
  et~al.}{2018}]{2018arXiv180909177M}
{Mondrik} N.,  {Newton} E.,   {Irwin} D.~C.~J.,  2018, preprint, \href
  {http://adsabs.harvard.edu/abs/2018arXiv180909177M} {} (\mn@eprint {arXiv}
  {1809.09177})

\bibitem[\protect\citeauthoryear{{Mullan} \& {Bais}}{{Mullan} \&
  {Bais}}{2018}]{2018ApJ...865..101M}
{Mullan} D.~J.,  {Bais} H.~P.,  2018, \mn@doi [\apj]
  {10.3847/1538-4357/aadfd1}, \href
  {http://adsabs.harvard.edu/abs/2018ApJ...865..101M} {865, 101}

\bibitem[\protect\citeauthoryear{{Mullan}, {MacDonald}, {Dieterich}  \&
  {Fausey}}{{Mullan} et~al.}{2018}]{2018arXiv181104149M}
{Mullan} D.~J.,  {MacDonald} J.,  {Dieterich} S.,   {Fausey} H.,  2018,
  preprint, \href {http://adsabs.harvard.edu/abs/2018arXiv181104149M} {}
  (\mn@eprint {arXiv} {1811.04149})

\bibitem[\protect\citeauthoryear{{Newton}, {Irwin}, {Charbonneau},
  {Berta-Thompson}, {Dittmann}  \& {West}}{{Newton}
  et~al.}{2016}]{2016ApJ...821...93N}
{Newton} E.~R.,  {Irwin} J.,  {Charbonneau} D.,  {Berta-Thompson} Z.~K.,
  {Dittmann} J.~A.,   {West} A.~A.,  2016, \mn@doi [\apj]
  {10.3847/0004-637X/821/2/93}, \href
  {http://adsabs.harvard.edu/abs/2016ApJ...821...93N} {821, 93}

\bibitem[\protect\citeauthoryear{Oliphant}{Oliphant}{2015}]{Oliphant:2015:GN:2886196}
Oliphant T.~E.,  2015, Guide to NumPy, 2nd edn.
CreateSpace Independent Publishing Platform, USA

\bibitem[\protect\citeauthoryear{{Osten} \& {Wolk}}{{Osten} \&
  {Wolk}}{2015}]{2015ApJ...809...79O}
{Osten} R.~A.,  {Wolk} S.~J.,  2015, \mn@doi [\apj]
  {10.1088/0004-637X/809/1/79}, \href
  {http://adsabs.harvard.edu/abs/2015ApJ...809...79O} {809, 79}

\bibitem[\protect\citeauthoryear{{Osten}, {Kowalski}, {Sahu}  \&
  {Hawley}}{{Osten} et~al.}{2012}]{2012ApJ...754....4O}
{Osten} R.~A.,  {Kowalski} A.,  {Sahu} K.,   {Hawley} S.~L.,  2012, \mn@doi
  [\apj] {10.1088/0004-637X/754/1/4}, \href
  {http://adsabs.harvard.edu/abs/2012ApJ...754....4O} {754, 4}

\bibitem[\protect\citeauthoryear{{Osten} et~al.,}{{Osten}
  et~al.}{2016}]{2016ApJ...832..174O}
{Osten} R.~A.,  et~al., 2016, \mn@doi [\apj] {10.3847/0004-637X/832/2/174},
  \href {http://adsabs.harvard.edu/abs/2016ApJ...832..174O} {832, 174}

\bibitem[\protect\citeauthoryear{{Owen} \& {Mohanty}}{{Owen} \&
  {Mohanty}}{2016}]{2016MNRAS.459.4088O}
{Owen} J.~E.,  {Mohanty} S.,  2016, \mn@doi [\mnras] {10.1093/mnras/stw959},
  \href {http://adsabs.harvard.edu/abs/2016MNRAS.459.4088O} {459, 4088}

\bibitem[\protect\citeauthoryear{{Pascucci} et~al.,}{{Pascucci}
  et~al.}{2011}]{2011ASPC..448..469P}
{Pascucci} I.,  et~al., 2011, in {Johns-Krull} C.,  {Browning} M.~K.,   {West}
  A.~A.,  eds,  Astronomical Society of the Pacific Conference Series Vol. 448,
  16th Cambridge Workshop on Cool Stars, Stellar Systems, and the Sun. p.~469
  (\mn@eprint {arXiv} {1101.1913})

\bibitem[\protect\citeauthoryear{{Paudel}, {Gizis}, {Mullan}, {Schmidt},
  {Burgasser}, {Williams}  \& {Berger}}{{Paudel}
  et~al.}{2018a}]{2018ApJ...858...55P}
{Paudel} R.~R.,  {Gizis} J.~E.,  {Mullan} D.~J.,  {Schmidt} S.~J.,  {Burgasser}
  A.~J.,  {Williams} P.~K.~G.,   {Berger} E.,  2018a, \mn@doi [\apj]
  {10.3847/1538-4357/aab8fe}, \href
  {http://adsabs.harvard.edu/abs/2018ApJ...858...55P} {858, 55}

\bibitem[\protect\citeauthoryear{{Paudel}, {Gizis}, {Mullan}, {Schmidt},
  {Burgasser}, {Williams}  \& {Berger}}{{Paudel}
  et~al.}{2018b}]{2018ApJ...861...76P}
{Paudel} R.~R.,  {Gizis} J.~E.,  {Mullan} D.~J.,  {Schmidt} S.~J.,  {Burgasser}
  A.~J.,  {Williams} P.~K.~G.,   {Berger} E.,  2018b, \mn@doi [\apj]
  {10.3847/1538-4357/aac8e0}, \href
  {http://adsabs.harvard.edu/abs/2018ApJ...861...76P} {861, 76}

\bibitem[\protect\citeauthoryear{{Pecaut} \& {Mamajek}}{{Pecaut} \&
  {Mamajek}}{2013}]{2013ApJS..208....9P}
{Pecaut} M.~J.,  {Mamajek} E.~E.,  2013, \mn@doi [\apjs]
  {10.1088/0067-0049/208/1/9}, \href
  {http://adsabs.harvard.edu/abs/2013ApJS..208....9P} {208, 9}

\bibitem[\protect\citeauthoryear{{Perez} \& {Granger}}{{Perez} \&
  {Granger}}{2007}]{2007CSE.....9c..21P}
{Perez} F.,  {Granger} B.~E.,  2007, \mn@doi [Computing in Science and
  Engineering] {10.1109/MCSE.2007.53}, \href
  {http://adsabs.harvard.edu/abs/2007CSE.....9c..21P} {9, 21}

\bibitem[\protect\citeauthoryear{{Ranjan}, {Wordsworth}  \&
  {Sasselov}}{{Ranjan} et~al.}{2017}]{2017ApJ...843..110R}
{Ranjan} S.,  {Wordsworth} R.,   {Sasselov} D.~D.,  2017, \mn@doi [\apj]
  {10.3847/1538-4357/aa773e}, \href
  {http://adsabs.harvard.edu/abs/2017ApJ...843..110R} {843, 110}

\bibitem[\protect\citeauthoryear{{Reiners} \& {Basri}}{{Reiners} \&
  {Basri}}{2009}]{2009ApJ...705.1416R}
{Reiners} A.,  {Basri} G.,  2009, \mn@doi [\apj]
  {10.1088/0004-637X/705/2/1416}, \href
  {http://adsabs.harvard.edu/abs/2009ApJ...705.1416R} {705, 1416}

\bibitem[\protect\citeauthoryear{{Reiners} et~al.,}{{Reiners}
  et~al.}{2018}]{2018A&A...612A..49R}
{Reiners} A.,  et~al., 2018, \mn@doi [\aap] {10.1051/0004-6361/201732054},
  \href {http://adsabs.harvard.edu/abs/2018A%26A...612A..49R} {612, A49}

\bibitem[\protect\citeauthoryear{{Ribas} et~al.,}{{Ribas}
  et~al.}{2016}]{2016A&A...596A.111R}
{Ribas} I.,  et~al., 2016, \mn@doi [\aap] {10.1051/0004-6361/201629576}, \href
  {http://adsabs.harvard.edu/abs/2016A%26A...596A.111R} {596, A111}

\bibitem[\protect\citeauthoryear{{Ricker}}{{Ricker}}{2014}]{2014JAVSO..42..234R}
{Ricker} G.~R.,  2014, Journal of the American Association of Variable Star
  Observers (JAAVSO), \href {http://adsabs.harvard.edu/abs/2014JAVSO..42..234R}
  {42, 234}

\bibitem[\protect\citeauthoryear{{Schmidt} et~al.,}{{Schmidt}
  et~al.}{2014}]{2014ApJ...781L..24S}
{Schmidt} S.~J.,  et~al., 2014, \mn@doi [\apjl] {10.1088/2041-8205/781/2/L24},
  \href {http://adsabs.harvard.edu/abs/2014ApJ...781L..24S} {781, L24}

\bibitem[\protect\citeauthoryear{{Schmidt}, {Hawley}, {West}, {Bochanski},
  {Davenport}, {Ge}  \& {Schneider}}{{Schmidt}
  et~al.}{2015}]{2015AJ....149..158S}
{Schmidt} S.~J.,  {Hawley} S.~L.,  {West} A.~A.,  {Bochanski} J.~J.,
  {Davenport} J.~R.~A.,  {Ge} J.,   {Schneider} D.~P.,  2015, \mn@doi [\aj]
  {10.1088/0004-6256/149/5/158}, \href
  {http://adsabs.harvard.edu/abs/2015AJ....149..158S} {149, 158}

\bibitem[\protect\citeauthoryear{{Segura}, {Walkowicz}, {Meadows}, {Kasting}
  \& {Hawley}}{{Segura} et~al.}{2010}]{2010AsBio..10..751S}
{Segura} A.,  {Walkowicz} L.~M.,  {Meadows} V.,  {Kasting} J.,   {Hawley} S.,
  2010, \mn@doi [Astrobiology] {10.1089/ast.2009.0376}, \href
  {http://adsabs.harvard.edu/abs/2010AsBio..10..751S} {10, 751}

\bibitem[\protect\citeauthoryear{{Shulyak}, {Reiners}, {Engeln}, {Malo},
  {Yadav}, {Morin}  \& {Kochukhov}}{{Shulyak}
  et~al.}{2017}]{2017NatAs...1E.184S}
{Shulyak} D.,  {Reiners} A.,  {Engeln} A.,  {Malo} L.,  {Yadav} R.,  {Morin}
  J.,   {Kochukhov} O.,  2017, \mn@doi [Nature Astronomy]
  {10.1038/s41550-017-0184}, \href
  {http://adsabs.harvard.edu/abs/2017NatAs...1E.184S} {1, 0184}

\bibitem[\protect\citeauthoryear{{Theissen}}{{Theissen}}{2018}]{2018ApJ...862..173T}
{Theissen} C.~A.,  2018, \mn@doi [\apj] {10.3847/1538-4357/aaccfa}, \href
  {http://adsabs.harvard.edu/abs/2018ApJ...862..173T} {862, 173}

\bibitem[\protect\citeauthoryear{{Tilley}, {Segura}, {Meadows}, {Hawley}  \&
  {Davenport}}{{Tilley} et~al.}{2017}]{2017arXiv171108484T}
{Tilley} M.~A.,  {Segura} A.,  {Meadows} V.~S.,  {Hawley} S.,   {Davenport} J.,
   2017, preprint, \href {http://adsabs.harvard.edu/abs/2017arXiv171108484T} {}
  (\mn@eprint {arXiv} {1711.08484})

\bibitem[\protect\citeauthoryear{{Tonry} et~al.,}{{Tonry}
  et~al.}{2012}]{2012ApJ...750...99T}
{Tonry} J.~L.,  et~al., 2012, \mn@doi [\apj] {10.1088/0004-637X/750/2/99},
  \href {http://adsabs.harvard.edu/abs/2012ApJ...750...99T} {750, 99}

\bibitem[\protect\citeauthoryear{{Van Grootel} et~al.,}{{Van Grootel}
  et~al.}{2018}]{2018ApJ...853...30V}
{Van Grootel} V.,  et~al., 2018, \mn@doi [\apj] {10.3847/1538-4357/aaa023},
  \href {http://adsabs.harvard.edu/abs/2018ApJ...853...30V} {853, 30}

\bibitem[\protect\citeauthoryear{{Vida}, {K{\H o}v{\'a}ri}, {P{\'a}l},
  {Ol{\'a}h}  \& {Kriskovics}}{{Vida} et~al.}{2017}]{2017ApJ...841..124V}
{Vida} K.,  {K{\H o}v{\'a}ri} Z.,  {P{\'a}l} A.,  {Ol{\'a}h} K.,   {Kriskovics}
  L.,  2017, \mn@doi [\apj] {10.3847/1538-4357/aa6f05}, \href
  {http://adsabs.harvard.edu/abs/2017ApJ...841..124V} {841, 124}

\bibitem[\protect\citeauthoryear{Vin{\'\i}cius, Barentsen, Hedges,
  Gully-Santiago  \& Cody}{Vin{\'\i}cius et~al.}{2018}]{lightkurve}
Vin{\'\i}cius Z.,  Barentsen G.,  Hedges C.,  Gully-Santiago M.,   Cody A.~M.,
  2018, KeplerGO/lightkurve, \mn@doi{10.5281/zenodo.1181928}, \url
  {http://doi.org/10.5281/zenodo.1181928}

\bibitem[\protect\citeauthoryear{{West} et~al.,}{{West}
  et~al.}{2011}]{2011AJ....141...97W}
{West} A.~A.,  et~al., 2011, \mn@doi [\aj] {10.1088/0004-6256/141/3/97}, \href
  {http://adsabs.harvard.edu/abs/2011AJ....141...97W} {141, 97}

\bibitem[\protect\citeauthoryear{{Williams} \& {Cieza}}{{Williams} \&
  {Cieza}}{2011}]{2011ARA&A..49...67W}
{Williams} J.~P.,  {Cieza} L.~A.,  2011, \mn@doi [\araa]
  {10.1146/annurev-astro-081710-102548}, \href
  {http://adsabs.harvard.edu/abs/2011ARA%26A..49...67W} {49, 67}

\bibitem[\protect\citeauthoryear{{Yashiro}, {Akiyama}, {Gopalswamy}  \&
  {Howard}}{{Yashiro} et~al.}{2006}]{2006ApJ...650L.143Y}
{Yashiro} S.,  {Akiyama} S.,  {Gopalswamy} N.,   {Howard} R.~A.,  2006, \mn@doi
  [\apjl] {10.1086/508876}, \href
  {http://adsabs.harvard.edu/abs/2006ApJ...650L.143Y} {650, L143}

\bibitem[\protect\citeauthoryear{{Youngblood} et~al.,}{{Youngblood}
  et~al.}{2017}]{2017ApJ...843...31Y}
{Youngblood} A.,  et~al., 2017, \mn@doi [\apj] {10.3847/1538-4357/aa76dd},
  \href {http://adsabs.harvard.edu/abs/2017ApJ...843...31Y} {843, 31}

\makeatother
\end{thebibliography}
%\begin{thebibliography}{99}
%\end{thebibliography}

%%%%%%%%%%%%%%%%%%%%%%%%%%%%%%%%%%%%%%%%%%%%%%%%%%

% Don't change these lines
\bsp	% typesetting comment
\label{lastpage}
\end{document}